\documentclass[intlimits,twoside,a4paper]{article}

\usepackage[T2A]{fontenc} 
\usepackage[utf8]{inputenc} 
\usepackage[greek,ukrainian,english]{babel}
\usepackage{cmpj3}

\usepackage{amsmath}
\usepackage{empheq}
\usepackage{pstricks}
\usepackage{mathrsfs}
\usepackage{mathtools}
\usepackage{cancel}
\usepackage{accents}
\usepackage{dsfont}
\usepackage{color}
\usepackage[normalem]{ulem}
\usepackage[caption = false]{subfig}

\usepackage{teubner}

\articletype{A part of the Special collection to the 100th anniversary of the birth of Ihor Yukhnovskii}

\issue{2025}{28}{2}{23401}
\doinumber{10.5488/CMP.28.23401}

\title[Effective and asymptotic scaling in a 1D billiard problem]{Effective and asymptotic scaling in a one-dimensional billiard problem}

\author[T. Holovatch, Yu. Kozitsky, K. Pilorz, Yu. Holovatch]{
			{T. Holovatch}\orcid{0009-0005-2953-8730} \refaddr{inst1,inst2}, 
			{Yu. Kozitsky}\orcid{0000-0002-4320-8835}\refaddr{inst3}, 
			{K. Pilorz}\orcid{0000-0003-2596-3260}\refaddr{inst3}, 
			{Yu. Holovatch}\orcid{0000-0002-1125-2532}\refaddr{inst1,inst2,inst4,inst5}
    }
\addresses{
		\addr{inst1}{Institute for Condensed Matter Physics of the National Academy of Sciences of Ukraine, 79011 Lviv, Ukraine}
   \addr{inst2}{{$\mathbb{L}^4$ Collaboration and Doctoral College for the Statistical Physics of Complex Systems}, 
			{Lviv-Leipzig-Lorraine-Coventry}, 
			{Europe}}
		\addr{inst3}{{Institute of Mathematics, Maria Curie-Sk\l odowska University}, 
			20-031 Lublin,
			{Poland}}
		\addr{inst4}{{Centre for Fluid and Complex Systems, Coventry University}, 
			Coventry CV1 5FB,
			{UK}}
		\addr{inst5}{Complexity Science Hub Vienna, 1030 Vienna, Austria} 
		}

\date{Received March 26, 2025}

\begin{document}
\maketitle
\begin{abstract}
 The emergence of power
  laws that govern the large-time dynamics of a one-dimensional billiard of $N$ point particles is analysed.
In the initial state, the resting particles are placed in the positive half-line $x\geqslant 0$ at equal distances.   
Their masses alternate between two distinct values. The dynamics is initialized by giving the leftmost particle
a positive velocity. Due to elastic inter-particle collisions, the 
  whole system gradually comes into motion, filling both right-hand and left-hand half-lines.	As shown
  by [{Chakraborti S., Dhar A., Krapivsky P., SciPost Phys., 2022, \textbf{13}, 074}],
an inherent feature of such a billiard is the emergence of two different modes: the shock wave 
that propagates in  $x\geqslant 0$ and the splash region  in  $x<0$. Moreover, the behaviour of the relevant
observables  is characterized by universal asymptotic power-law dependencies. 
In view of the finite size of the system and of finite observation times, these dependencies only start to 
acquire a universal character. To analyse them, we set up molecular dynamics simulations using the concept of effective scaling exponents, 
familiar in the theory of continuous phase transitions. We present results for the effective exponents
that govern the large-time behaviour of the shock-wave front, the number of collisions, the energies and 
momentum of different modes and analyse their tendency to approach corresponding universal values.
      \keywords 
		billiards, one-dimensional cold gas, shock wave, scaling, scaling exponents, molecular dynamics 
		\end{abstract}

\hspace{17em}\textit{{To the 100th anniversary of Professor Ihor Yukhnovskii}} 
\vspace{2ex}

In the year of centenary of the outstanding Ukrainian public and political figure Ihor Yukhnovskii (September 1, 1925 -- March 26, 2024), the scientific community also celebrates him as the founder of Lviv school of statistical physics.
Two of the authors of this article (Yu.K. and Yu.H.) had the honor of being among Prof. Yukhnovskii's direct students 
and working with him in the field of phase transition theory.  His major contributions in this area include the
pioneering work on the non-perturbative approach to the analysis of critical behaviour at continuous phase transitions, 
as summarized later in one of his monographs \cite{Yukhnovskii87}. The topic we discuss in this article --- the behaviour 
of a one-dimensional multi-particle billiard --- may at first glance seem to be far removed from the problem of phase transitions. 
What is common, however, is the phenomenon of scaling --- the emergence of universal power-law asymptotics --- that govern the behaviour of different observables in both the continuous phase transition problem and in the billiards problem. This is what led us to the choice of 
this topic. After all, a deeper analysis of such an analogy may allow one to use methods and even concepts that are already well established in the study of one set of phenomena for the study of other problems. This is exactly the task we set in this work. 

The further structure of the article is as follows. In the next section \ref{I} we  describe 
the phenomenon we are interested in and the observables in terms of which it is customary to describe this 
phenomenon. In particular, we discuss the behaviour of a system of $N$ elastically interacting particles after one of 
them is given a certain velocity at an initial time $t$, the $N$-particle billiard problem. Theoretical 
analysis suggests that for a large number of particles  in such a system, a large-time power-law 
asymptotics of various observables is established. The scaling exponents that govern this asymptotics 
are universal: they are solely defined by space dimension and billiard symmetry. However,
for a  realistic case of finite $N$ and~$t$,  the effective exponents might not yet reach their 
asymptotic values. This observation evokes a natural analogy with effective critical exponents, which 
describe the power-law behaviour of the thermodynamic or structural characteristics of a system undergoing 
a continuous phase transition. Such exponents are introduced into phase transition theory when 
analysing finite-size systems or when describing critical behaviour beyond a transition point. 
Their introduction is discussed in section \ref{II}. In order to track an emergence of the
universal scaling laws, we set up the molecular dynamics simulations for a version of a 
one-dimensional billiard,  as described in section \ref{III}. We end with conclusions and 
outlook in section \ref{IV}.

\section{One-dimensional many-particle billiards: phenomenon and its characteristics}	\label{I}

In this paper, we consider the behaviour of the so-called multi-particle billiards.
Mathematically,  billiards describe the motion of mass points with elastic 
reflections \cite{Sinai76,Tabachnikov05}. 
Billiards is not a single mathematical theory, this is rather a  field to elaborate or
test various methods, both analytical and numerical.
Usually billiards are studied in the framework of dynamical systems theory
 and concern the behaviour of a single particle
 in a domain with elastic 
 reflections from the boundary. A celebrated example
is given by the Sinai billiard, as an illustration for interacting Hamiltonian system 
that displays physical thermodynamical properties \cite{Sinai76}. One of the reasons for 
considering the problem of multi-particle billiards is that in this case
they can serve as an
intermediate bridge to explain the processes in a continuous medium based on 
elementary acts of inter-particle interaction via rigorous deviation of continuous 
hydrodynamics description starting from the atomistic one. 
This ambitious task is also known as Hilbert's 6th problem, suggested by
David Hilbert in 1900 as ``\ldots{} the problem of developing mathematically 
the limiting processes \ldots{} which lead from the atomistic view to the laws of motion
of continua\ldots{}'', see, e.g., \cite{Gorban18} for more discussion.

\begin{figure}[htbp]
	\centerline{\includegraphics[angle=0,origin=c,width=0.75\linewidth]{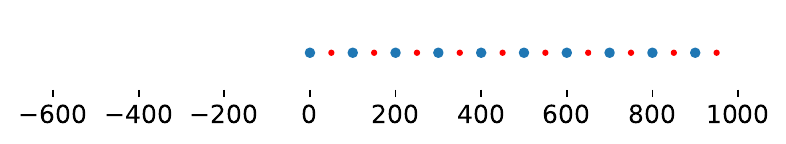}}
	\centerline{\textbf{(a)}}
	\centerline{\includegraphics[angle=0,origin=c,width=0.75\linewidth]{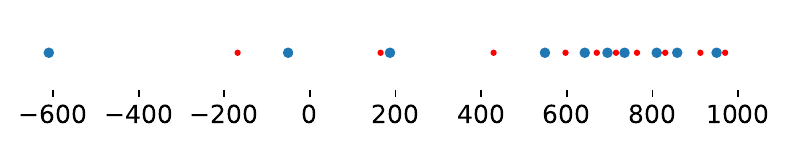}}
	\centerline{   \textbf{(b)}}
	
	\caption{(Colour online)
		The phenomenon we are interested in this article. $N$ point-like particles 
		of possibly different masses are at rest and form a one-dimensional chain, that fills in
		the half-space $x\geqslant 0$, panel \textbf{(a)}. 
		The first particle acquires certain initial velocity, starts moving to the right and, thus, due to 
		elastic collisions, the whole system gradually comes into motion, filling both right-hand and left-hand subspaces,
		panel \textbf{(b)}.	An inherent feature of the dynamics of such a billiard is the emergence of 
		two different modes~\cite{Chakraborti22}: the shock wave that propagates in the right-hand side 
		$x\geqslant 0$ and the splash region formed  by the particles that move 
		with negative velocities in the left-hand side $x<0$.		
		In our analysis we consider particles of two masses $M$ and $m$ 
		(blue and red balls) assuming that they alternate and are equidistantly located at the 
		initial moment of time.
	 }
	\label{fig1}
\end{figure}

We  focus on the behaviour of one-dimensional billiards.
In particular, we  consider the set of $N$ point-like particles, as 
depicted in figure~\ref{fig1}. In the initial state, all particles --- `billiard balls' ---
are at rest in the half-space $x\geqslant 0$ and are fixed at certain locations,  figure~\ref{fig1}a. Then, the first 
ball acquires an initial velocity $v_0$ directed to the right in the figure.  It undergoes 
an elastic collision with the second ball, then the next collision occurs, and gradually,
under the influence of elastic inter-ball collisions, that conserve momentum and energy, 
the whole system begins to move,  figure~\ref{fig1}b. We are interested in the 
asymptotic behaviour of such a billiard, which arises in the limit of a large number of 
balls and large observation times. 

One immediately notes that the problem is trivial 
when all the particles are of the same mass. Indeed, since the collisions are elastic, 
after the first particle is set in motion, a domino effect is observed: the first particle
takes the place of the second one and stops, the second particle takes the place of 
the third, and so on.  This degeneracy is removed when particle masses are different.
In our analysis we  consider particles of two masses $M$ and $m$ assuming that
they alternate, as shown by blue and red balls in figure~\ref{fig1}.
Such a system of alternating particles is known as an alternating hard particle gas. It 
is widely used for analysis of heat transport and hydrodynamics in one dimension
\cite{Garrido01,Dhar01,Grassberger02}, for the recent studies see~\cite{Lepri20,Chakraborti22} and references therein.

We should also  note the intrinsic symmetry of this problem: while at the 
initial moment of time, the right-hand half-space $x\geqslant 0$ is filled with particles, 
the left-hand half-space $x< 0$ is empty. This distinguishes it from a similar problem 
where particles fill the entire space and the dynamics is initiated by letting
to move the  particle at the origin. This latter case was considered in reference~\cite{Antal08}, where it was shown that the system large-time dynamics is governed 
by power laws. In particular, for the $d$-dimensional billiard, the number of
collisions ${\cal C}(t)$ and the number of moving particles ${\cal N}(t)$ were
found to scale as ${\cal C}(t)\sim t^\eta$, ${\cal N}(t)\sim t^\xi$ with $\eta=2(d+1)/(d+2)$,
 $\xi =2d/(d+2)$, leading to $\eta= 2\xi = 4/3$ for $d=1$. Moreover, noticing
 that the release of energy, when the initial particle comes to motion,
makes the phenomenon similar to the explosion in a continuous medium, it is
argued that the above scaling laws coincide with those that describe
 the shock  wave emanating from an explosion in the hydrodynamic theory \cite{Landau87}.
In turn, it leads to the power-law asymptotics for the shock-wave front
${\cal R}(t)\sim t^\delta$, with $\delta=2/(d+2)$. The above-mentioned symmetry 
restriction of the billiard of figure~\ref{fig1} leads to a number of unusual 
surprising phenomena. A detailed analysis of the dynamics 
of such a billiard showed \cite{Chakraborti22} that its asymptotic behaviour is characterized by two regimes:
the shock wave that propagates in the right-hand side $x\geqslant 0$ and the splash region
formed in the left-hand side $x<0$ by the particles that move with negative velocities.
The shock-wave front is described by the hydrodynamic equations, although differing 
from those of the symmetric case \cite{Antal08,Chakraborti21,Ganapa21}. Notably,
the particles in the splash region $x<0$ after experiencing a period of collisions  
move ballistically  and their movement  is non-hydrodynamic.

Let us quote some of the asymptotic laws that govern the dynamics of the one-dimensional
splash problem as they were derived in reference~\cite{Chakraborti22}:
\begin{itemize}
	\item With the span of time, all energy is accumulated in the splash region $x<0$.
	 Energy of particles in the right-hand half-space decays as
	 \begin{equation}\label{1}
	 	{\cal E}_{\rm{x}\geqslant 0}( t ) \sim t^{-\beta}\, .
	 \end{equation}  
 \item The shock-wave front (i.e., the coordinate of the front-right moving particle)
 propagates sub-ballistically with the exponent $\delta<1$:
  \begin{equation}\label{2}
 	{\cal R}( t ) \sim t^{\delta}\, .
 \end{equation} 
 \item The number of the inter-particle collisions grows as
  \begin{equation}\label{3}
 {\cal C}(t)\sim t^\eta \, .
 \end{equation}
 \item The growth of total momenta of particles in the left-hand and in the right-hand half-spaces, 
 ${\cal P}_{\rm{x}<0} ( t )$ and  ${\cal P}_{\rm{x}\geqslant 0} ( t )$ is governed by the same 
 exponent:
  \begin{equation}\label{4}
 	-\, {\cal P}_{\rm{x}<0} ( t )\sim {\cal P}_{\rm{x} \geqslant 0} ( t ) \sim  t^\gamma \, .
 \end{equation}
Obviously, the sum ${\cal P}_{\rm{x} \geqslant 0} ( t ) + {\cal P}_{\rm{x}<0} ( t )$ is constant and
due  to the momentum conservation law it equals the momentum of the incipient moving particle.
\end{itemize}  

\begin{table}[htb]
	\caption{Asymptotic values of the exponents governing the scaling laws (\ref{1})--(\ref{4})
		for the one-dimensional splash problem
		\cite{Chakraborti22}.
		\vspace{0.7em}
		\label{tab1}}
	\begin{center}
		\tabcolsep1.2mm
		\begin{tabular}{|c|c|c|c|}
			\hline
			$\beta$ & $\delta$ & $\eta$ & $\gamma$  \\			
			\hline
			0.11614383675 & 0.6279520544 & 1.255904109 & 0.2559041088 \\
			\hline 
		\end{tabular}
	\end{center}
\end{table}

Along with these asymptotic laws, the relations 
between the exponents governing the power-law behaviour 
equations~(\ref{1})--(\ref{4}) were derived:
\begin{equation}\label{5}
	\delta=\frac{2-\beta}{3}\, , \hspace{2em} \eta= 2\,\delta\, , \hspace{2em} \gamma = \frac{1-2\beta}{3}\, ,
\end{equation}
and the values of the exponents were obtained \cite{Chakraborti22}. The latter are displayed in table 
\ref{tab1}. 
Needless to emphasize the striking similarity of the relations (\ref{5}) with the familiar scaling relations
arising in the theory of critical phenomena \cite{Fisher67,Stanley87,Berche22}.

Strictly speaking, universal power-law asymptotics, equations~(\ref{1})--(\ref{4}) with the exponent values
given in table \ref{tab1} are observed in the limit of an infinite number of particles
(or in the hydrodynamic limit of a continuous medium). It is natural to expect (as we will see below this is indeed the case) that the laws are violated for finite observation times and system sizes. 
 Again, the situation is similar to the scaling laws governing the critical behaviour at continuous phase
 transitions. There, the universal asymptotics is inherent only in the thermodynamic limit and
 at the transition point, otherwise non-universal exponents are observed. 
In order to use this analogy further, in the next section we show how the asymptotic (universal) and the effective (system-specific) critical exponents are introduced in the phase transition description.

\section{Asymptotic and effective  exponents in the continuous phase transition description}	\label{II}

The concept of a critical exponent is one of the central concepts in the theory of continuous 
phase transitions. Critical exponents describe the singular power-law behaviour at a phase transition. They are universal (common to a wide class of different systems) and are the subject of a detailed analysis 
in the experiment, in theory, and in computer simulations \cite{Fisher67,Stanley87,Berche22}.
Strictly speaking, the asymptotic critical exponents are universal,  which are determined directly at the 
phase transition point. For definiteness, let us take a thermodynamic phase transition as an example, when the controlling parameter is temperature~$T$, and the other parameters, such as the external field, 
have already acquired their critical values. Then, the \textit{asymptotic critical exponent} $\upsilon$ for the 
observable  ${\cal O}(\tau)$ is defined as the limit:
\begin{equation}\label{6}
	\upsilon = - \lim_{\tau\to 0}\frac{\ln {\cal O}(\tau)}{\ln\tau} \, ,
\end{equation}
where $\tau=|T-T_c|/T_c$ is the dimensionless temperature measuring the distance to the critical point $T_c$.
Formula (\ref{6}) means that the leading singularity for ${\cal O}(\tau)$ is of the power-law form:
\begin{equation}\label{7}
{\cal O}(\tau) \sim \tau^{-\upsilon} (1+\dots)\, , \hspace{2em} \tau \to 0\, ,
\end{equation}
and the dots denote the sub-leading terms. 

As mentioned above, the asymptotic critical exponents (\ref{6}), (\ref{7}) are universal. They depend only on such global 
factors as space dimension, inter-particle interaction range, system symmetry. However, in practice, both in 
experiments and in computer simulations, one has to deal with non-universal critical 
exponents, which depend on the system details. These are commonly called \textit{effective critical exponents}.
At least two reasons for the appearance of such an effective critical behaviour can be pointed out. These are: 
(i) measurements outside the critical point $T_c$ (i.e., at $\tau\neq 0$) and (ii) simulations of finite-size systems.
Indeed, in an experiment, whether computer or real, it is not the singularity itself that is observed, 
but only the tendency for its appearance. To be more specific concerning these reasons, herein below we refer to the usual ways the effective exponents are introduced:
\begin{itemize}
	\item (i)
To describe the critical behaviour when the asymptotic scaling (\ref{7}) is not yet reached,
it is convenient to define the effective critical exponents via \cite{Kouvel64,Riedel74}:
\begin{equation}\label{8}
	\upsilon_{\rm eff}(\tau)=-\frac{{\rm d}\ln {\cal O}(\tau)}{{\rm d}\ln\tau}\, .
\end{equation}
Obviously,  the effective exponents coincide with the
asymptotic ones in the limit $\tau \to 0$.

\item (ii)
To study the emergence of power-law  singularities by computer simulations with a finite number of particles $N$, 
the ideas of finite-size scaling, FSS,  apply \cite{Privman90,Ardourel23}. 
They are based on the fact that when approaching $\tau=0$, the correlation length $\xi$ becomes the only 
characteristic scale of the system. At $\tau=0$, the correlation length diverges with the critical
exponent $\nu$, $\xi\sim \tau^{-\nu}$. In turn, inverting this dependence as $\tau \sim \xi^{-1/\nu}$, 
enables one to represent the power-law asymptotic~(\ref{7}) as a dependence of the observable on 
the correlation length: ${\cal O} \sim \xi^{\upsilon/\nu}$. 
Assuming that the maximal correlation length increases linearly with the system size $L$, one 
arrives\footnote{This assumption
holds only below the so-called upper critical dimension $d_{\rm uc}$. For $d>d_{\rm uc}$, the correlation-length scaling is 
governed by the critical exponent \textqoppa \,\,(koppa): $\xi \sim L^{\text{\qoppa}}$, \textqoppa~$=d/d_{\rm uc}$. However, this discussion goes beyond the scope of this paper~\cite{Berche22}. 
 }
at the FSS relation, determining the leading power-law singularity as:
\begin{equation}\label{9}
	{\cal O}(L) \sim L^{\upsilon/\nu} (1+\dots)\, , \hspace{2em} L \to \infty\, ,
\end{equation}
and, as in equation~(\ref{8}) the dots denote the sub-leading terms. Since computer simulations 
are carried out for finite $L$, in practice the effective exponents are determined:
\begin{equation}\label{10}
	\frac{\upsilon}{\nu}|_{\rm eff}(L)=\frac{{\rm d}\ln {\cal O}(L)}{{\rm d}\ln L}\, .
\end{equation}
\end{itemize}
It is worth noting that reasons (i)  and (ii) are also related 
by the fact that  the correlation length $\xi$ cannot be infinite for a finite-size system. The same 
applies to an infinite system beyond the critical point $\tau=0$.

Effective critical exponents and their evolution to
the asymptotic values are the subject of detailed studies, 
see, for example, \cite{Folk06,Dudka03,Perumal03} or recent papers 
\cite{Ruiz-Lorenzo22,Ruiz-Lorenzo25} for more references. In the mentioned papers, 
effective critical exponents describe continuous phase transitions in various 
magnetic, structurally disordered, or fluid systems. 
In the next section, we apply this concept to the analysis of power-law 
scaling in the one-dimensional billiard splash problem. 

\section{Molecular dynamic simulations of a one-dimensional $N$-particle billiard}	\label{III}

In this section, we present the results of computer simulations of the one-dimensional
$N$-particle billiard shown in figure~\ref{fig1}.  Note that, unlike the simulations of 
one-dimensional billiards in references~\cite{Antal08,Chakraborti21,Ganapa21,Chakraborti22}, where the initial coordinates of 
particles were considered as random, in our case  at time $t=0$, the particles are located at equal 
distances, as shown in figure~\ref{fig1}a.  Thus, there is no need to perform the averaging over different starting configurations of particle locations. 
We performed the molecular dynamics (MD) simulations, solving the equations of motion 
for a system of $N=10\,000$ particles. We took the particles to be of two different
masses, $M$ and $m$ (blue and red balls in figure~\ref{1}), assuming that they are 
point-like and are placed at equal distances $r_0$ at the initial moment of time $t=0$. We 
consider that the incipient particle is of a larger mass $M>m$ and it acquires,  at $t=0$,
the initial velocity $v_0$ directed to the right. 
Since all particles subsequently interact through elastic collisions,  the total 
momentum and total energy of all particles are conserved throughout 
the entire evolution of the system and are equal to the initial energy and momentum of 
the incipient particle, $E_0=Mv_0^2/2$ and  $P_0=Mv_0$, correspondingly.
Note that the evolution of the billiard in which the first particle is of lighter mass
$m$ is analogous. Indeed, after a collision with the second heavier particle, the 
first particle acquires a negative velocity and continues its collisionless\footnote{Although
	this statement sounds reasonable, it has been proven only under certain conditions
	and still remains as a conjecture~\cite{Chakraborti22}.}  leftward 
motion at a constant velocity. Instead, the subsequent dynamics of the system is 
initiated by the heavier particle of mass $M$, which moves rightward, as in the previous case.

\begin{figure}[htbp]
	\centerline{ 
		\includegraphics[angle=0,origin=c,width=0.5\linewidth]{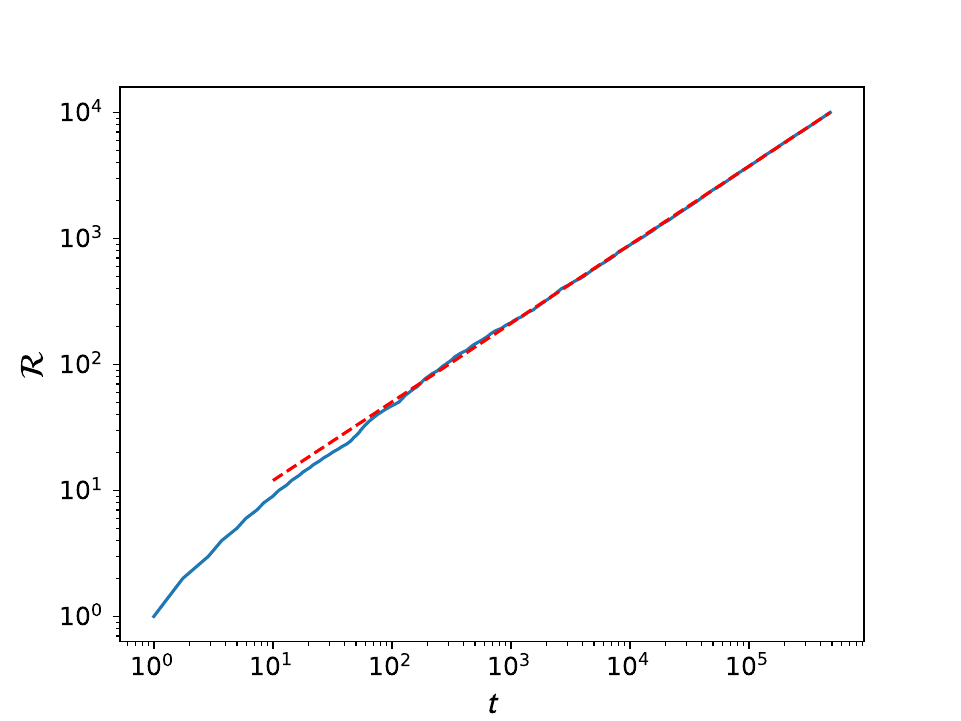}
		\includegraphics[angle=0,origin=c,width=0.5\linewidth]{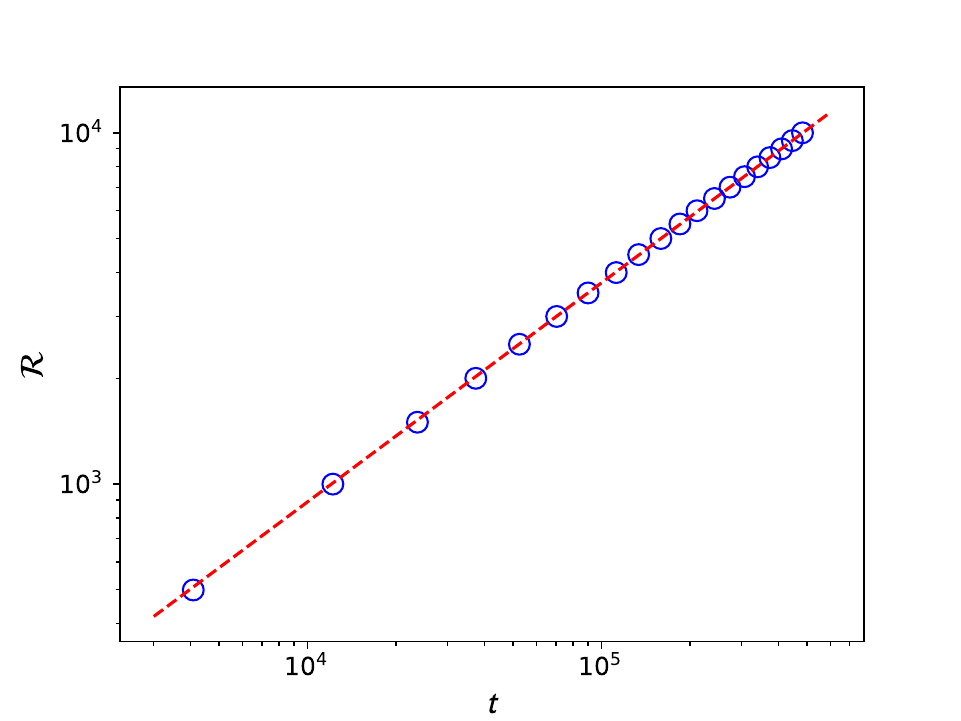}	}
	\centerline{\textbf{(a)} \hspace{21em} \textbf{(b)}}
	\centerline{	\includegraphics[angle=0,origin=c,width=0.5\linewidth]{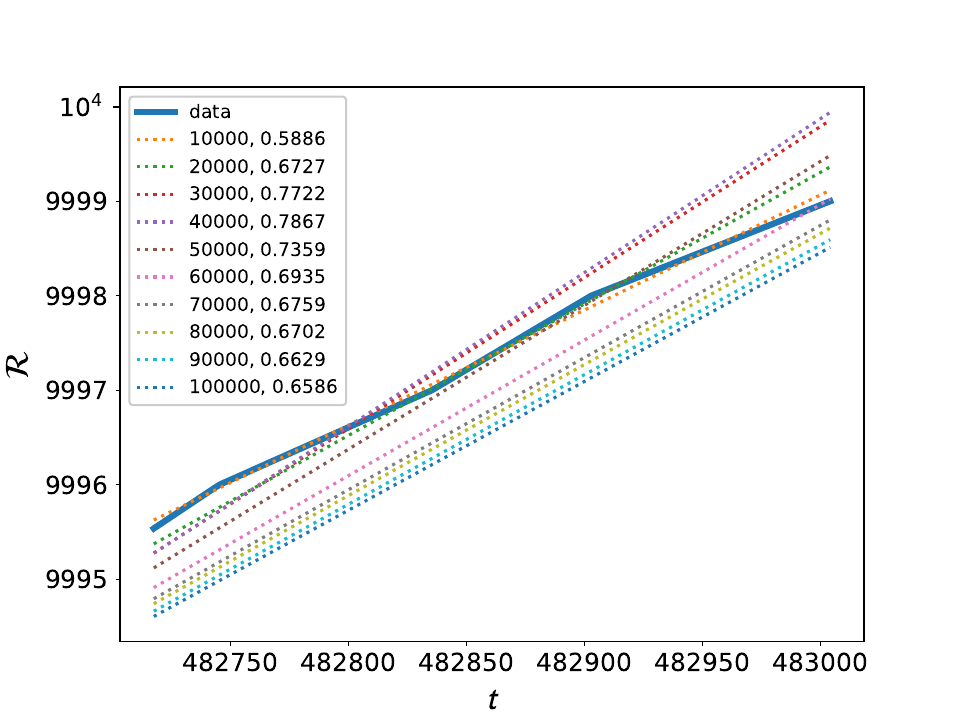}	\includegraphics[angle=0,origin=c,width=0.5\linewidth]{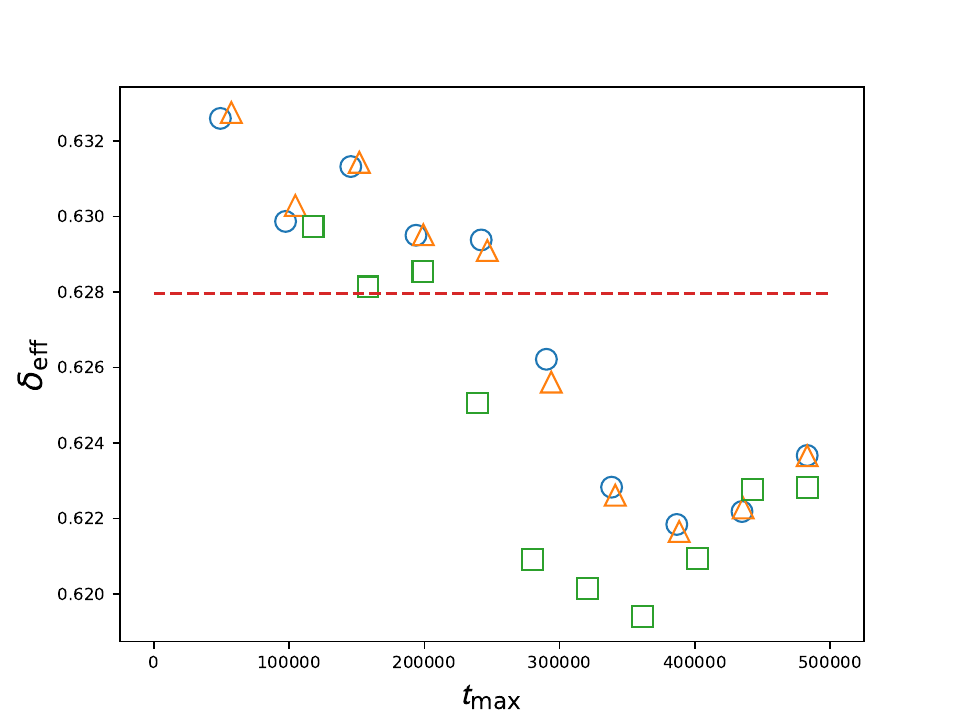}}
	\centerline{\textbf{(c)} \hspace{21em} \textbf{(d)}}
	
	\caption{(Colour online)
		Shock-wave front position ${\cal R}(t)$ for the one-dimensional billiard of particles of 
		two alternating masses, $M/m=2$,  that are 
		equidistantly located at the initial moment of time, blue and red balls in figure~\ref{fig1}.
		\textbf{(a):} Data points (seen as a blue solid curve) and their fit
		by the power-law dependency $R(t)\sim t^{\delta}$ (red dashed line). \textbf{(b):} Data points (blue circles) correspond to the time instances when the shock wave passes each $500\times k$th particle, $k=1,\dots ,20$. Red dashed line is a guideline for the eye and shows a power-law dependency  with the asymptotic exponent $\delta=0.62795$, see table \ref{tab1}. 
		\textbf{(c):} Data points
		(seen as a blue solid curve) correspond to the time interval when the last $10^4$ collisions took place. The dashed lines describe the fitted power-law dependencies taking into account the last $n=10^4\times k$ collisions, $k=1,\dots,10$. The legend displays $n$ together
		with the exponent $\delta_{\rm eff}$ as it results from the fit for each $n$.  
		\textbf{(d):}  Effective exponent $\delta_{\rm eff}$ obtained as a result of a power-law fit for ${\cal R}(t)$ in the region 
		$[t_{\rm min}, t_{\rm max}]$ with  $t_{\rm min}= 10^3,\,10^4,\,t_{10\%}$, shown by circles, triangles, and 
		squares, correspondingly,  and  
		$t_{10\%}=117\,753$ is the time when 10\% of all collisions have occurred.
		In each case, the maximal value of $ t_{\rm max}$ is limited by the time when the shock wave 
		reaches the rightmost particle $t_{\rm max}=  t_{\rm final} = 483\,004$.}
	\label{fig2}
\end{figure}

We are interested in the time evolution of the shock-wave front
${\cal R}( t )$, the total energy  of particles  
in the region $x\geqslant 0$,  ${\cal E}_{\rm{x}\geqslant 0}( t )$, the 
number of collisions ${\cal C}(t)$, and the total momenta of particles  
in the splash region $x<0$,  ${\cal P}_{\rm{x}<0} ( t )$.
In particular, our goal is to check whether their time dynamics are characterized by 
power-law asymptotics (\ref{1})--(\ref{4}) with the exponents given in table \ref{tab1}.
In what follows, we present the results in dimensionless variables, measuring distance, 
energy, momentum, and time in units of $r_0$, $E_0$, $P_0$, and $t_0=r_0/v_0$, 
respectively.  Since we are dealing with a finite number of particles, the observation time 
is limited by time $t_{\rm final}$, when the last of $N=10\,000$ particles is set in motion.
For the parameters chosen in our simulations ($E_0=1$, $M/m=2$), the last, rightmost, particle 
starts moving after the system experiences $C_{\rm final}=13\,369\,302$  collisions. This corresponds 
to $t_{\rm final}=483\,004$ in the dimensionless variables. Accordingly, all further results  
concern the time window between $t=0$ and $t=t_{\rm final}= 483\,004$.

Blue curve in figure~\ref{fig2}a shows the time evolution of the shock-wave front position 
(coordinate of the rightmost moving particle) ${\cal R}(t)$.
After rather short initialization period, $t\simeq 10^2 - 10^3$, the system tends to develop
a power-law dependence ${\cal R}(t)\sim t^{\delta}$, as shown by the dashed red line in the 
double logarithmic plot in the figure. The latter
is displayed more in detail in figure~\ref{fig2}b, where the blue circles mark the coordinates 
of the shock-wave front as it reaches every 500th particle. The red dashed line in the figure shows 
the power-law dependence with the asymptotic value of the exponent $\delta$ taken from table \ref{tab1}.
At first glance, such a coincidence of simulated and analytically predicted data can serve as 
a good confirmation that the asymptotic scaling is observed.
However, the linear dependence depicted in this figure serves only as an eye-guide,
since it is not obtained as a result of the simulation data fit. In fact, the situation is more 
complicated and typical of the analysis of finite-sized systems.
Figure~\ref{fig2}c shows the behaviour of the tail of ${\cal R}(t)$: its change in time is
always discrete (blue curve) and the value of the power-law fit exponent $\delta_{\rm eff}$ will depend on the 
time window within which such a fit is performed (dashed lines). The legend shows the values of the time
windows together with the values of the effective exponents that correspond to the fits.
It is obvious that the time windows chosen for the illustration in figure~\ref{fig2}c  are too 
small to judge about the asymptotics. However, the general problem remains: the time 
intervals, on the one hand, should correspond to large times (starting from some 
`equilibration' time $t_{\rm min}$ when the system has experienced a comparatively large number of collisions) and, on the other hand, the time window should be large 
enough  to smoothen the fluctuations, that are inevitably present even for large times.
With these two caveats in mind, we approach finding the effective scaling exponent
$\delta_{\rm eff}$ as depicted in figure~\ref{fig2}d. The exponent $\delta_{\rm eff}$ 
shown in this figure corresponds to power-law fits for different time intervals $[t_{\rm min},t_{\rm max}]$.
We made  three different choices for $t_{\rm min}$: $t_{\rm min}= 10^3,\,10^4,\,t_{10\%}$, shown by circles, triangles, and 
squares, correspondingly. Time $t_{10\%}=117\,753$ is the time when 10\% of all collisions have occurred.
In turn, for each $t_{\rm min}$, the value of $t_{\rm max}$ increased up to $t_{\rm final}$, when the shock wave 
reached the rightmost particle. The dependence of $\delta_{\rm eff}(t_{\rm max })$ with an increase of $t_{\rm max}$ 
is shown in the figure. As one can see from the figure, with an increase of $t_{\rm max}$, $\delta_{\rm eff}$ 
tends to approach its asymptotic value shown by the red dashed line. This trend is maintained for all considered initial times
$t_{\rm min}$.

\begin{figure}[htbp]

\centerline{	
	\includegraphics[angle=0,origin=c,width=0.5\linewidth]{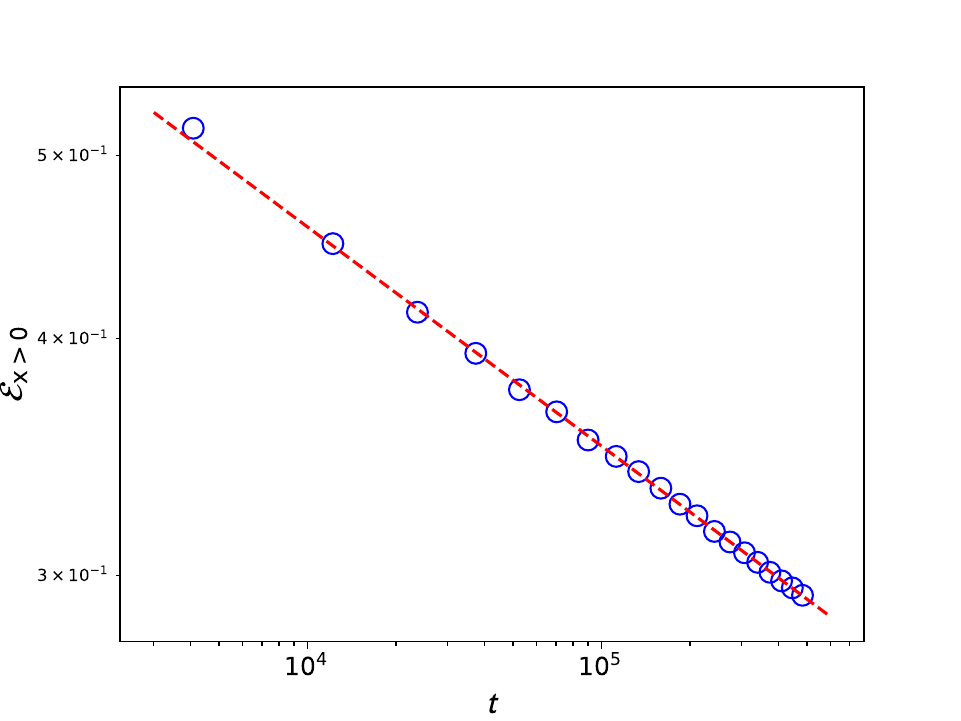}
\includegraphics[angle=0,origin=c,width=0.5\linewidth]{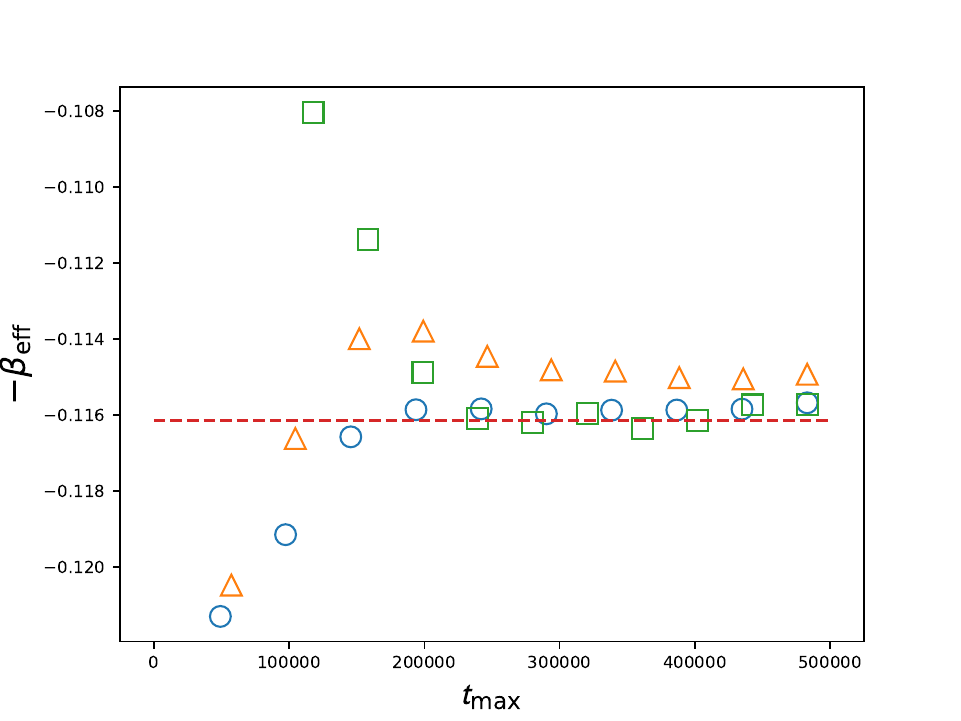}
}
	\centerline{\textbf{(a)} \hspace{21em} \textbf{(b)}}

\centerline{
	\includegraphics[angle=0,origin=c,width=0.5\linewidth]{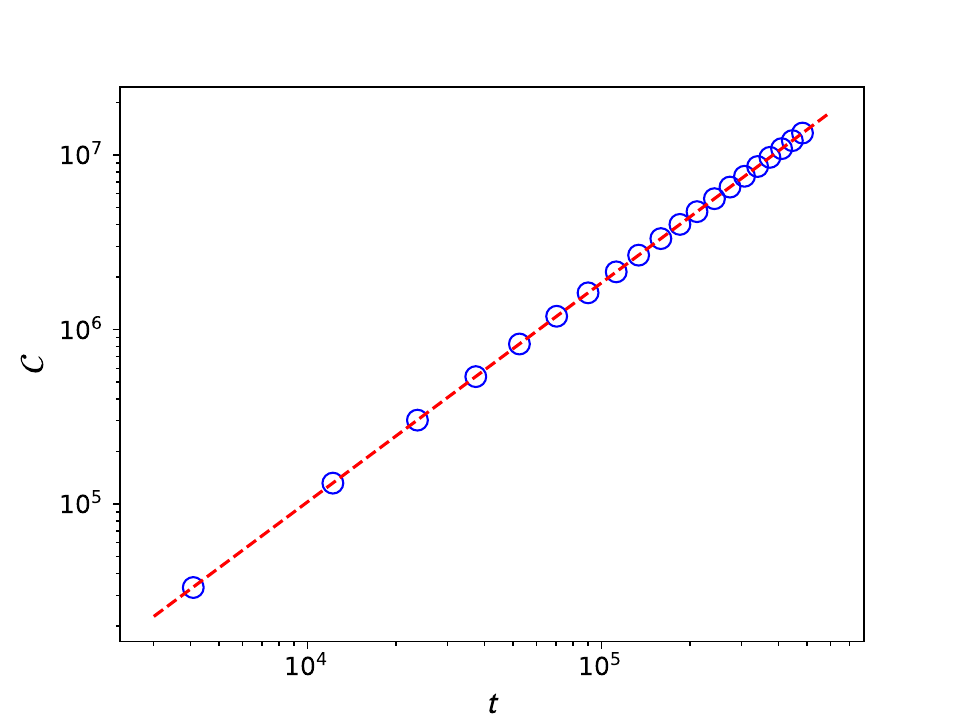}
	\includegraphics[angle=0,origin=c,width=0.5\linewidth]{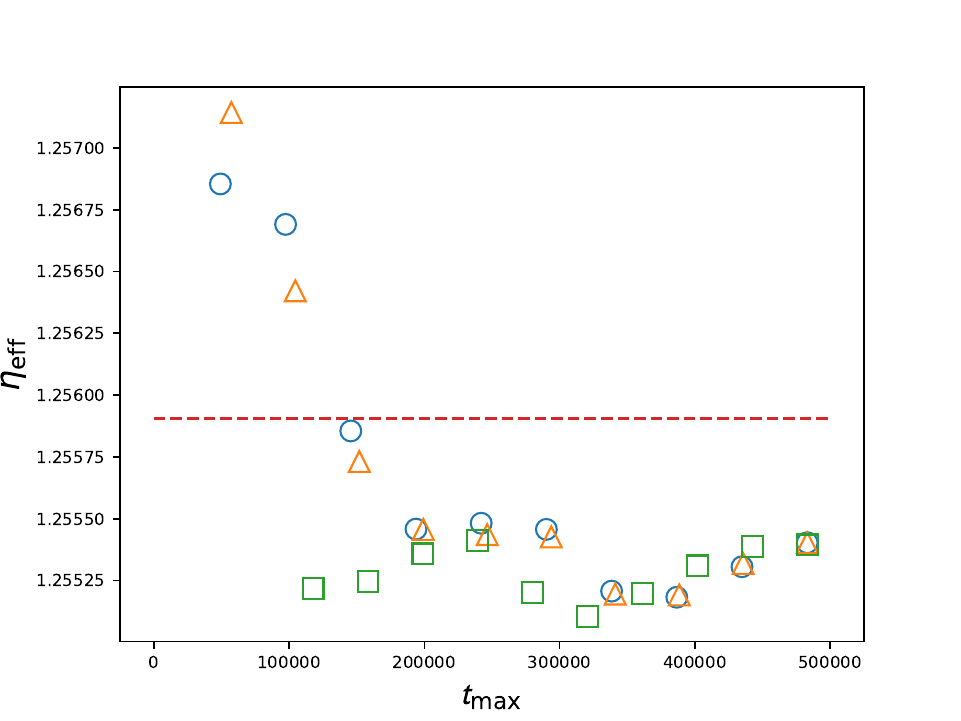}
}
\centerline{\textbf{(c)} \hspace{21em} \textbf{(d)}}

\centerline{
	\includegraphics[angle=0,origin=c,width=0.5\linewidth]{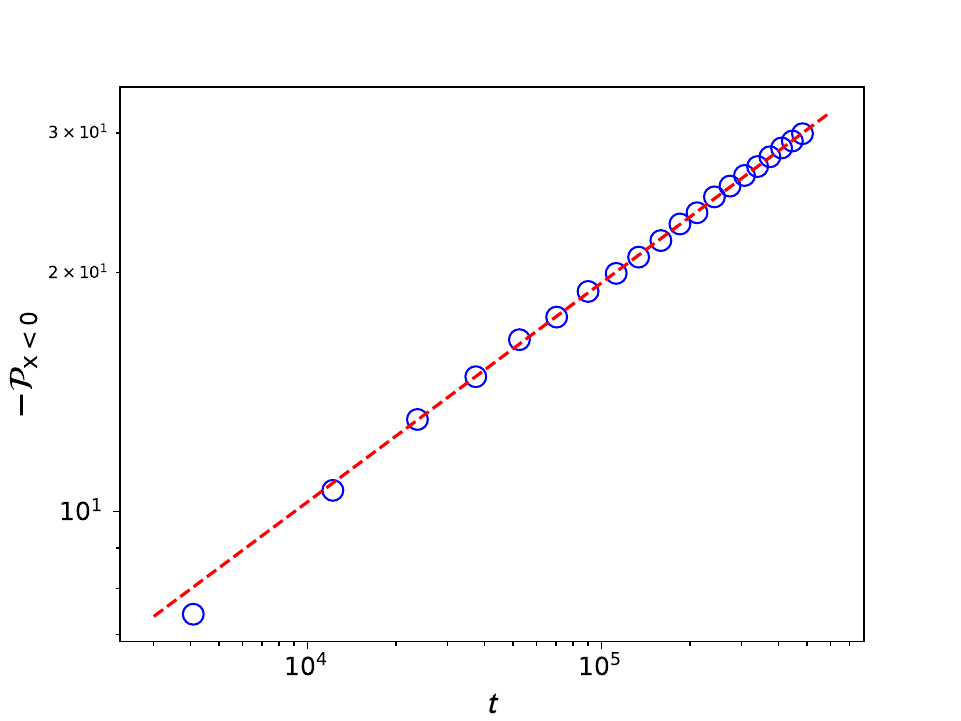}
\includegraphics[angle=0,origin=c,width=0.5\linewidth]{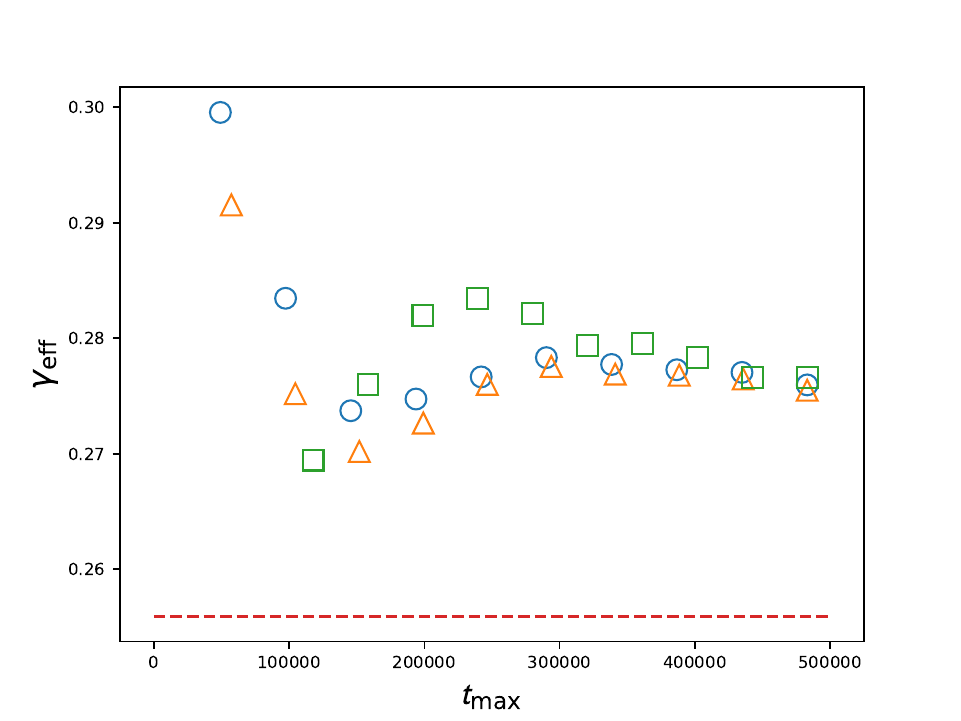}
}
\centerline{\textbf{(e)} \hspace{21em} \textbf{(f)}}

	\caption{(Colour online) \textit{Left panels} display time dependencies of \textbf{(a):} total energy of 
		particles  in the $x \geqslant 0$ region ${\cal E}_{\rm{x} \geqslant 0}( t )$,  
		\textbf{(c):} number of collisions ${\cal C}(t)$, and \textbf{(e):} total momentum of particles in the $x < 0$ region
		${\cal P}_{\rm{x}<0} ( t )$. Data points (blue circles) correspond to time instances when the shock wave 
		passes each $500\times k$th particle, $k=1,\dots ,20$. Red dashed lines show power-law dependencies  with 
		the asymptotic exponents, as given in table \ref{tab1}. \textit{Right panels} give the effective scaling exponents
		for the quantities displayed in the left panels. Similar as in figure~\ref{fig2}d, the exponents 
		were obtained  as a result of power-law fits in the region 
		$[t_{\rm min}, t_{\rm max}]$ with  $t_{\rm min}= 10^3,\,10^4,\, t_{10\%}$, shown by circles, triangles, and 
		squares, correspondingly. See the text and the caption of figure~\ref{fig2}d for more description. Red dashed lines show
		values of the asymptotic exponents given in Table \ref{tab1}. 
	}
	\label{fig3}
\end{figure}

Based on the results of the shock-wave front ${\cal R}(t)$ analysis, we defined in a similar way the effective 
scaling exponents for  the 
total energy of  particles in the $x \geqslant 0$ region ${\cal E}_{\rm{x} \geqslant 0}( t )$, the total number of collisions ${\cal C}(t)$,   and the total momentum of particles in the $x < 0$ region ${\cal P}_{\rm{x}<0} ( t )$.  The results are shown in figure~\ref{fig3}.
Similar to figure~\ref{fig2}b, the plots in figures~\ref{fig3}a, \ref{fig3}c, \ref{fig3}e exhibit power-law 
behaviour, which at first glance is well described by the asymptotic exponents in table~\ref{tab1}. 
However, a more detailed analysis shows the scaling with varying effective exponents, as depicted in the plots
of the right-hand column of figure~\ref{fig3}.
With an increasing observation time, all exponents show a tendency to approach their asymptotic values. However, 
the convergence is different. 
While the effective exponents $\delta_{\rm eff}$, $\beta_{\rm eff}$, $\eta_{\rm eff}$ differ from their
asymptotic counterparts within the third significant digit only, the exponent $\gamma_{\rm eff}$
for the total momentum of 
particles in the $x < 0$ region ${\cal P}_{\rm{x}<0} ( t )$ remains further away, $\gamma_{\rm eff} - \gamma \simeq 0.02$.

\section{Conclusions and outlook}	\label{IV}

A characteristic feature of the billiard problem we have considered here is the lack of a priori
randomness, neither in the distribution of masses nor in 
the inter-particle distances. Therefore, the emergence of the hydrodynamic power-law 
asymptotics --- pointing to the stochastic background of the underlying process ---  
may be interpreted as a kind of self-averaging in the system.  As it follows from the simulations we have performed, the exponents
that govern the scaling of different observables are system-dependent. In our case, 
such a dependence manifests itself as their change with the  system size $N$
and the size of the time window in which the fit is performed. 
To interpret such a behaviour, it is instructive to make use of the concept of 
effective critical exponents, as introduced in the theory of continuous phase 
transitions. The observed effective critical exponents, figures~\ref{fig2}d, \ref{fig3})b,
\ref{fig3})d, \ref{fig3})f, tend to change with the growth of $N$ (therefore,  with the growth of the time window too) and approach their asymptotic values displayed in table \ref{tab1}. 
As mentioned above, these values were obtained analytically by solving
equations of a continuous medium. Therefore, this trend also serves as an indirect indication
that the considered here discrete system of equidistantly distributed alternating hard particles 
reproduces the continuous medium in the limiting case.

\section*{Acknowledgement}

We thank Pavel Krapivsky, whose talk at the Ising lectures in Lviv \cite{Ising_lectures} motivated 
us to perform this study.


\ukrainianpart

\title{Ефективний та асимптотичний скейлінґ в одній задачі про одновимірний більярд}
\author{{Т. Головач}\refaddr{inst1,inst2}, {Ю. Козицький}\refaddr{inst3}, {К. Пілож}\refaddr{inst3}, {Ю. Головач}\refaddr{inst1,inst2,inst4,inst5}}
\addresses{
	\addr{l1}{Інститут фізики конденсованих систем НАН України, 79011 Львів, Україна}
	\addr{l2}{Співпраця $\mathbb{L}^4$ і Коледж докторантів ``Статистична фізика складних систем'', Ляйпціґ-Лотаринґія-Львів-Ковентрі, Європа}
	\addr{l3}{Інститут математики, Університет Марії Склодовської-Кюрі,  Люблін, 20-031, Польща}
	\addr{l4}{Центр плинних і складних систем, Університет Ковентрі, Ковентрі, CV1 5FB, Велика Британія}
	\addr{l5}{Центр науки про складність, Відень, 1030, Австрія}
}

\makeukrtitle

\begin{abstract} 
	Проаналізовано появу степеневих законів, які керують довгочасовою 
	динамікою одновимірного більярду $N$ точкових частинок. У початковому стані 
	частинки розташовані в додатній півлінії $x\geqslant 0$ на однаковій відстані. 
	Їхні маси чергуються між двома різними значеннями. 
	Динаміка ініціалізується наданням крайній лівій частинці додатної швидкості. 
	Завдяки пружним міжчастинковим зіткненням вся система поступово приходить в рух, 
	заповнюючи як праву, так і ліву півлінії. 
	Як показано в [{Chakraborti~S., Dhar~A., Krapivsky~P., SciPost Phys., 2022, \textbf{13}, 074}],
	особливістю такого більярду є поява двох різних режимів: ударної хвилі
	що поширюється в $x\geqslant 0$ та області бризок в $x<0$.
	 Крім того, поведінка відповідних спостережуваних характеризується універсальними асимптотичними степеневими залежностями. З огляду на скінчений розмір системи і скінчений час спостереження ці залежності тільки починають набувати універсального характеру. Для їх аналізу ми проводимо симуляції методом молекулярної динаміки і використовуємо відому в теорії неперервних фазових переходів концепцію ефективних показників скейлінґу. Ми представляємо результати для ефективних показників, які керують довгочасовою поведінкою фронту ударної хвилі, кількості зіткнень, енергії та імпульсу різних мод та аналізуємо їх тенденцію наближатися до відповідних універсальних значень.
	
	\keywords 
	більярд, одновимірний холодний газ, ударна хвиля, скейлінґ, показники скейлінґу, молекулярна динаміка 
\end{abstract}

\end{document}